\documentstyle[aps,epsf]{revtex}

\newcommand{\be}{\begin{equation}}
\newcommand{\ee}{\end{equation}}
\newcommand{\bea}{\begin{eqnarray}}
\newcommand{\eea}{\end{eqnarray}}

\newcommand{\aeq}{&=&}

\newcommand{\itDelta}{{\it \Delta}}

\newcommand{\itPi}{{\it \Pi}}

\newcommand{\bra}{\langle}
\newcommand{\ket}{\rangle}
\newcommand{\dbra}{\bra \! \bra}
\newcommand{\dket}{\ket \! \ket}

\newcommand{\me}{\mbox{e}}

\newcommand{\bq}{{\bar q}}

\newcommand{\rT}{{\rm T}}

\newcommand{\rR}{{\rm R}}

\newcommand{\rRe}{{\rm Re}}

\newcommand{\rN}{{\rm N}}
\newcommand{\rS}{{\rm S}}

\begin{document}

%%%%%%%%%%%%%%%%%%%%%%%%%%%%%%%%%%%%%%%%%%%%%%%%%%%%%%%
% Title etc.
%%%%%%%%%%%%%%%%%%%%%%%%%%%%%%%%%%%%%%%%%%%%%%%%%%%%%%%

\draft

\title{Analysis of Velocity Fluctuation in Turbulence based on 
Generalized Statistics
}

\author{T.~Arimitsu
%\thanks{arimitsu@cm.ph.tsukuba.ac.jp}
}
\address{Institute of Physics, University of Tsukuba,
Ibaraki 305-8571, Japan 
}
\author{N.~Arimitsu
%\thanks{arimitsu@dnj.ynu.ac.jp}
}
\address{Graduate School of EIS, Yokohama Nat'l.\ University,
Kanagawa 240-8501, Japan 
}

\date{10 October 2001}

\maketitle

\begin{abstract}

The numerical experiments of turbulence conducted by Gotoh et al.\ are analyzed 
precisely with the help of the formulae for the scaling exponents of 
velocity structure function and for the probability density function of 
velocity fluctuations.
These formulae are derived by the present authors with the multifractal aspect 
based on the statistics that are constructed on the generalized measures of 
entropy, i.e., the extensive R\'{e}nyi's or the non-extensive Tsallis' entropy. 
It is shown, explicitly, that there exist {\it two} scaling regions, i.e., 
the {\it upper} scaling region with larger separations which may correspond 
to the scaling range observed by Gotoh et al.,
and the {\it lower} scaling region with smaller separations which is 
a new scaling region extracted first by the present systematic analysis.
These scaling regions are divided by a definite length approximately of 
the order of the Taylor microscale, which may correspond to the crossover length
proposed by Gotoh et al.\ as the low end of the scaling range 
(i.e., the {\it upper} scaling region).
%It indicates that the multifractal distribution of singularities 
%in velocity gradient in turbulent flow is robust enough to produce scaling
%behaviors even for the phenomena outside the inertial range.

\end{abstract}


%\narrowtext

%%%%%%%%%%%%%%%%%%%%%%%%%%%%%%%%%%%%%%%%%%%%%%%%%%%%%%%%%%%%%%%%%%%
% Text
%%%%%%%%%%%%%%%%%%%%%%%%%%%%%%%%%%%%%%%%%%%%%%%%%%%%%%%%%%%%%%%%%%%

%\section{Introduction}

After the discovery of the Kolmogorov spectrum \cite{K41},
there have been plenty of trials including \cite{Heisenberg48,Landau59}
in order to understand the intermittent evolution of fluid 
in fully developed turbulence.
Among them, the line based on a kind of {\it ensemble} theoretical approaches, 
initiated by the log-normal model \cite{Oboukhov62,K62,Yaglom},
consists of the $\beta$-model (a uni-fractal dimensional analysis) \cite{Frisch78},
the p-model (a multifractal model) \cite{Meneveau87a,Meneveau87b},
the 3D binomial Cantor set model \cite{Hosokawa91} and so on.
On this line, an investigation of turbulence based on the generalized entropy, 
i.e., R\'{e}nyi's \cite{Renyi} or Tsallis' \cite{Tsallis88,Tsallis99},
was started by the present authors~\cite{AA,AA1,AA2,AA3,AA4,AA5,AA6}.
After a rather preliminary investigation of the p-model \cite{AA},
it has been developed further to derive the analytical expression for 
the scaling exponents of velocity structure function \cite{AA1,AA2,AA3,AA4}, and to
determine the probability density function (PDF) of velocity fluctuations 
\cite{AA4,AA5,AA6} by a self-consistent statistical mechanical approach.

With the help of the analytical formulae derived in \cite{AA1,AA2,AA3,AA4,AA5,AA6},
we will analyze, in this paper, the PDF's of velocity fluctuations observed
in the beautiful DNS (i.e., the direct numerical simulation)
conducted by Gotoh et al.\ \cite{Gotoh01}.
We will deal with the data at the Taylor microscale Reynolds number 
$R_\lambda = 381$, since at this Reynolds number 
Gotoh et al.\ observed the PDF with accuracy up to order of 
$10^{-9} \sim 10^{-10}$, in contrast with any previous experiments, real or numerical.
We showed in \cite{AA5} that our formulae can explain quite well 
the PDF's observed in the real experiment by Lewis and Swinney 
\cite{Lewis-Swinney99} for turbulent Couette-Taylor flow
at $R_\lambda = 270$ $(\rRe = 5.4 \times 10^{5})$
produced in a concentric cylinder system in which, however, 
the PDF's were measured only with accuracy of order of 10$^{-5}$.
Note that the success of the present theory in the analysis of 
this turbulent Couette-Taylor flow
may indicate the robustness of singularities associated with velocity gradient
even for the case of no inertial range.
We have already made clear in \cite{AA6} that the present theory can
also explain quite well the PDF's of {\it longitudinal} velocity fluctuations
reported by Gotoh et al.\ \cite{Gotoh01},
and have revealed the superiority of our PDF to the one derived in 
\cite{Beck-Lewis-Swinney01} when the accuracy is raised up to 
the order of 10$^{-9}$.
Assured by the success of these rather preliminary tests,
we will apply our theory for further precise analyses of the data obtained 
in \cite{Gotoh01} including the PDF's of {\it transverse} velocity fluctuations 
in addition to those of {\it longitudinal} fluctuations.

The basic equation describing fully developed turbulence is 
the Navier-Stokes equation
$
\partial {\vec u}/\partial t
+ ( {\vec u}\cdot {\vec \nabla} ) {\vec u} 
= - {\vec \nabla} \left(p/\rho \right)
+ \nu \nabla^2 {\vec u}
\label{N-S eq}
$
of an incompressible fluid, where $\rho$, $p$ and 
$\nu$ represent, respectively, the mass density, 
the pressure and the kinematic viscosity.
Our main interest is the correlation of 
measured time series of the streamwise velocity component $u$, 
say $x$-component, of the fluid velocity field ${\vec u}$
in the turbulent flow produced by a grid with size $\ell_0$
putting in a laminar flow.
Within Taylor's frozen flow hypothesis, 
the quantity of our interest reduces to
$
\delta u(r) = \vert u(x+r) - u(x) \vert
$
representing the spatial difference of the component $u$.
We assume that in the downstream of the grid there appears
a cascade of eddies with different sizes $\ell_n = \delta_n \ell_0$ 
where $\delta_n = \delta^{-n}$ $(\delta > 1,\ n=0,1,2,\cdots)$.
At each step of the cascade, say at the $n$th step, eddies break up into 
$\delta$ pieces producing an energy cascade with the energy-transfer rate
$\epsilon_n$ that represents the rate of transfer of energy per unit mass 
from eddies of size $\ell_n$ to those of size $\ell_{n+1}$.

The Reynolds number $\rRe$ of the system is given by 
$
{\rm Re} = \delta u_0 \ell_0/\nu = ( \ell_0/\eta )^{4/3}
$
where
$
\eta = ( \nu^3/\epsilon )^{1/4}
$
is the Kolmogorov scale~\cite{K41}
with the energy input rate $\epsilon$ to the largest eddies
of size $\ell_0$, i.e., $\epsilon_0 = \epsilon$.
Here, we introduced the notation
$
\delta u_n = \delta u(\ell_n)
$
representing the velocity difference across a distance $r \sim \ell_n$.
Then, our main interest in the following reduces to the fluctuation
of velocity difference $\delta u_n$ corresponding to 
the size of the $n$th eddies in the cascade.
Note that the dependence of the number of steps $n$ on $r/\eta$, 
within the analysis where intermittency is not taken into account \cite{K41}, 
is given by
\be
n = - \log_\delta r/\eta + (3/4) \log_\delta \rRe.
\label{n-roeta no intermittency}
\ee
For homogeneous and isotropic turbulence,
there is a relation between the Taylor microscale Reynolds number $R_\lambda$
and the Reynolds number $\rRe$ \cite{Batchelor53}:
$
R_\lambda^2 = A \rRe
$
where $A$ is a real number of the order of $0.01 \sim 10$ depending on 
the experimental setup.

For high Reynolds number $\rRe \gg 1$ or for the situation where 
effects of the kinematic viscosity can be neglected compared with
those of the turbulent viscosity, the Navier-Stokes equation
for incompressible fluid is invariant under 
the scale transformation~\cite{Frisch-Parisi83,Meneveau87b}:
${\vec r} \rightarrow \lambda^{\alpha/3} {\vec u}$, 
${\vec u} \rightarrow \lambda {\vec r}$, 
$t \rightarrow \lambda^{1- \alpha/3} t$ and 
$\left(p/\rho\right) \rightarrow \lambda^{2\alpha/3} \left(p/\rho\right)$.
Here, the exponent $\alpha$ is an arbitrary real quantity which specifies the degrees
of singularity in the velocity gradient
$
\left\vert \partial u(x)/\partial x \right\vert 
= \lim_{\ell_n \rightarrow 0} \vert u(x+\ell_n) - u(x) \vert/\ell_n
= \lim_{\ell_n \rightarrow 0} \delta u_n/\ell_n
$.
This can be seen with the relation
$
\delta u_n / \delta u_0 = (\ell_n / \ell_0)^{\alpha/3},
\label{u-alpha}
$
which leads to the singularity in the velocity gradient~\cite{Benzi84}
for $\alpha < 3$, since 
$
\delta u_n/\ell_n^{\alpha/3} = {\rm const.}
$.
We also have the relation 
$
\epsilon_n/\epsilon = \left(\ell_n/\ell_0 \right)^{\alpha -1}
\label{u/epsilon-alpha}
$.

Let us determine the probability 
$
P^{(n)}(\alpha) d\alpha
$
to find at a point in physical space an eddy of size $\ell_n$ which
has a value of the degree of singularities in the range 
$
\alpha \sim \alpha + d \alpha
$.
Assuming that each step in the cascade is statistically independent, i.e.,
$
P^{(n)}(\alpha) = [P^{(1)}(\alpha)]^n
$,
our task reduces to determine $P^{(1)}(\alpha)$.
In order to proceed, we have assumed \cite{AA1,AA2,AA3,AA4,AA5,AA6}
that the underlying statistics
describing the intermittent evolution of fully-developed turbulence is
the one given by the R\'{e}nyi entropy \cite{Renyi}
$
S_{q}^{\rR}[P^{(1)}(\alpha)] = \left(1-q \right)^{-1} 
\ln \int d \alpha P^{(1)}(\alpha)^{q}
\label{SqR-alpha}
$,
or by the Tsallis entropy \cite{Tsallis88,Tsallis99,Havrda-Charvat}
$
S_{q}^{\rT}[P^{(1)}(\alpha)] = \left(1-q \right)^{-1} 
\left(\int d \alpha P^{(1)}(\alpha)^{q} -1 \right)
\label{SqTHC-alpha}
$.
Taking an extremum of these entropies with appropriate constraints, 
i.e., the normalization of distribution function:
$
\int d\alpha P^{(1)}(\alpha) = \mbox{const.}
\label{cons of prob}
$
and the $q$-variance being kept constant as a known quantity:
$
\sigma_q^2 = \bra (\alpha- \alpha_0)^2 \ket_{q}
= (\int d\alpha P^{(1)}(\alpha)^{q} 
(\alpha- \alpha_0 )^2 ) / \int d\alpha P^{(1)}(\alpha)^{q}
\label{q-variance}
$,
we have \cite{AA1,AA2,AA3,AA4}
\be
P^{(1)}(\alpha) \propto \left[ 1 - (\alpha - \alpha_0)^2 \big/ (\itDelta \alpha )^2 
\right]^{1/(1-q)}
\label{Tsallis prob density}
\ee
with
$
(\itDelta \alpha)^2 = 2X \big/ [(1-q) \ln 2 ]
$.
The R\'{e}nyi entropy has an information theoretical basis, and has 
the extensive character as the usual thermodynamical entropy does.
On the other hand, the Tsallis entropy is non-extensive, and therefore
it provides us with one of the attractive trials for generalized statistical mechanics 
which deals with, for example, systems with long range correlations
with a hierarchical structure where the usual extensive characteristics 
in statistical mechanics are not applicable.
Note that the distribution functions which give an extremum of each entropy
have a common structure (see (\ref{Tsallis prob density}) for the present system
where $q \leq 1$)
in spite of different characteristics of these entropies, and
that the values of $\alpha$ are restricted within the range
$[\alpha_{\rm min},\ \alpha_{\rm max}]$, where
$\alpha_{\rm max} - \alpha_0 = \alpha_0 - \alpha_{\rm min} = \Delta \alpha$.
Note that
$
\sigma_q^2 = 2X/[(3-q) \ln 2]
$.
%we see that $\sigma_q^2$ is proportional to 
%the constraint we took when we derive the distribution function 
%(\ref{Tsallis prob density}) is consistent with the situation.
%We are informed by experimentalists 
%the value of the intermittency exponent $\mu$ that is proportional to
%$X$ in the region of $\mu$ where experiments are conducted.

By making use of an observed value of the intermittency exponent $\mu$ as an input,
the quantities $\alpha_0$, $X$ and the index $q$ can be determined, 
self-consistently, with the help of the three independent equations, i.e.,
the energy conservation:
$
\left\bra \epsilon_n \right\ket = \epsilon
\label{cons of energy}
$,
the definition of the intermittency exponent $\mu$:
$
\bra \epsilon_n^2 \ket 
= \epsilon^2 \delta_n^{-\mu}
\label{def of mu}
$,
and the scaling relation:
$
1/(1-q) = 1/\alpha_- - 1/\alpha_+
\label{scaling relation}
$
with $\alpha_\pm$ satisfying $f(\alpha_\pm) =0$ \cite{AA1,AA2,AA3,AA4}.
Here, the average $\bra \cdots \ket$ is taken with $P^{(n)}(\alpha)$.
The scaling relation is a generalization of the one derived first in
\cite{Costa,Lyra98} to the case where the multifractal spectrum has 
negative values.
For the region where the value of $\mu$ is usually observed,
i.e., $0.13 \leq \mu \leq 0.40$,
the three self-consistent equations are solved to give 
the approximate equations
$
\alpha_0 = 0.9989 + 0.5814 \mu
$,
$
X = - 2.848 \times 10^{-3} + 1.198 \mu
$
and
$
q = -1.507 + 20.58 \mu - 97.11 \mu^2 + 260.4 \mu^3 - 365.4 \mu^4 + 208.3 \mu^5
$.
These equations are slightly different from those given in \cite{AA5}, since 
the region of $\mu$ has been extended a little bit.

\begin{figure}[htbp]
\begin{center}
\leavevmode
\epsfxsize=100mm
\epsfbox{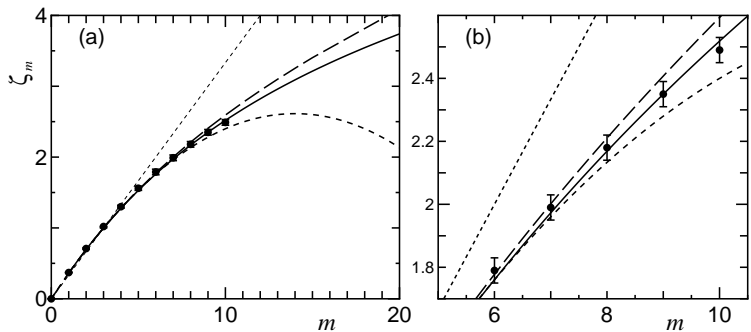}
\caption{The {\it longitudinal} scaling exponents $\zeta_m^{\rm L}$ 
of the velocity structure function.
The second figure (b) is a magnification of the first (a).
The results of DNS obtained by Gotoh et al.\ ($R_{\lambda} = 381$)
are shown by closed circles.
The present theoretical formula (\ref{zeta}) was used to determine
the value $\mu = 0.240$ of the intermittency exponent, and the result is
drawn by solid line.
Dotted line represents K41, whereas dashed line She-Leveque.
The prediction of the log-normal model is given 
by short-dashed line with $\mu = 0.240$.
}
\label{Fig: L scaling exponent}
\end{center}
\end{figure}

If we put $q=1$ from the beginning, i.e., starting with 
the Boltzmann-Shannon entropy, and take an extremum with the same
constraints given above for $q=1$, we have a Gaussian distribution function.
The parameters $\alpha_0$ and $X$ are then determined by two conditions,
i.e., energy conservation and the definition of the intermittency exponent. 
The obtained results \cite{AA4} completely coincide with those in
the log-normal model \cite{Oboukhov62,K62,Yaglom}. 
Therefore, the present approach can be interpreted as an extension of 
the one in the log-normal model.
%Note that log-normal model can be interpreted as a multifractal model.

\begin{figure}[thbp]
\begin{center}
\leavevmode
\epsfxsize=100mm
\epsfbox{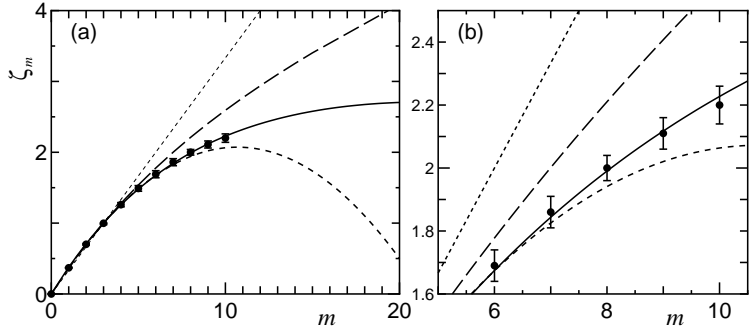}
\caption{The {\it transverse} scaling exponents $\zeta_m^{\rm T}$ 
of the velocity structure function.
The second figure (b) is a magnification of the first (a).
The results of DNS obtained by Gotoh et al.\ ($R_{\lambda} = 381$)
are shown by closed circles.
The present theoretical formula (\ref{zeta}) was used to determine
the value $\mu = 0.327$ of the intermittency exponent, and the result is
drawn by solid line.
Dotted line represents K41, whereas dashed line She-Leveque.
The prediction of the log-normal model is given 
by short-dashed line with $\mu = 0.327$.
}
\label{Fig: T scaling exponent}
\end{center}
\end{figure}

It is known that in turbulent flow there are two mechanisms to rule 
its dissipative evolution, i.e., the one controlled by the kinematic viscosity 
that takes care thermal fluctuations, and the other by the turbulent viscosity
that is responsible for intermittent fluctuations related to the singularities 
in velocity gradient.
Therefore, it may be reasonable to assume that the probability 
$\itPi^{(n)}(x_n) dx_n$
to find the scaled velocity fluctuation $\vert x_n \vert = \delta u_n/\delta u_0$ 
in the range $x_n \sim x_n+dx_n$ can be divided into two parts:
\be
\itPi^{(n)}(x_n) dx_n = \itPi_{\rN}^{(n)}(x_n) dx_n 
+ \itPi_{\rS}^{(n)}(\vert x_n \vert) dx_n.
\ee
Here, the normal part PDF $\itPi_{\rN}^{(n)}(x_n)$ stemmed from thermal dissipation,
and the singular part PDF $\itPi_{\rS}^{(n)}(\vert x_n \vert)$ 
from multifractal distribution of the singularities.
The latter is derived through
$
\itPi_{\rS}^{(n)}(\vert x_n \vert) dx_n = P^{(n)}(\alpha) d \alpha
$
with the transformation of the variables:
$
\vert x_n \vert = \delta_n^{\alpha/3}
$.

The $m$th moments of the velocity fluctuations, defined by
$
\dbra \vert x_n \vert^m \dket 
= \int_{-\infty}^{\infty} dx_n  
\vert x_n \vert^m \itPi^{(n)}(x_n)
$,
are given by 
\be
\dbra \vert x_n \vert^m \dket = 2 \gamma_m^{(n)} 
+ (1-2\gamma_0^{(n)} ) \
a_m \ \delta_n^{\zeta_m}
\ee
where
$
a_{3\bq} = \{ 2 / [C_{\bq}^{1/2} ( 1+ C_{\bq}^{1/2} ) ] \}^{1/2}
$
with
$
{C}_{\bq}= 1 + 2 \bq^2 (1-q) X \ln 2
\label{cal D}
$,
and
\be
2\gamma_m^{(n)} = \int_{-\infty}^{\infty} dx_n\ 
\vert x_n \vert^m \itPi_\rN^{(n)}(x_n).
\ee
We used the normalization: $\dbra 1 \dket = 1$.
The quantity 
\be
\zeta_m = \alpha_0 m /3 
- 2Xm^2 \big/\left[9 \left(1+{C}_{m/3}^{1/2} \right) \right] 
- \left[1-\log_2 \left(1+{C}_{m/3}^{1/2} \right) \right] 
\big/(1-q) 
\label{zeta}
\ee
is the so-called scaling exponent of the velocity structure function,
whose expression was derived first by the present authors \cite{AA1,AA2,AA3,AA4}.
In this paper, we will use the formula (\ref{zeta}) in order to extract 
the value of the intermittency exponent $\mu$ for the best fit to the measured 
scaling exponents by the method of least squares.

\begin{figure}[thbp]
\begin{center}
\leavevmode
\epsfxsize=70mm
\epsfbox{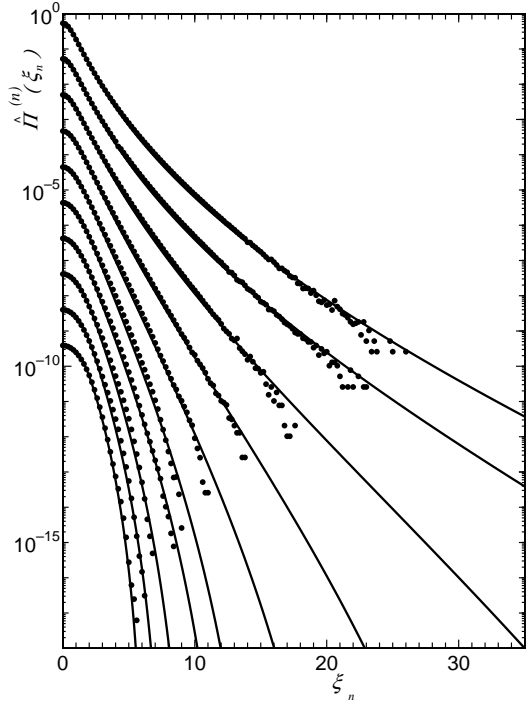}
\caption{Experimentally measured PDF of the {\it longitudinal} velocity fluctuations by 
Gotoh et al.\ for $R_\lambda =381$ 
are compared with the present theoretical results $\hat{\itPi}^{(n)}(\xi_n)$.
Closed circles are the symmetrized points  obtained by taking 
averages of the left and the right hand sides DNS data. 
For the experimental data, the separations $r/\eta = \ell_n/\eta$ 
are, from top to bottom: 
2.38, 4.76, 9.52, 19.0, 38.1, 76.2, 152, 305, 609 and 1220. 
Solid lines represent the curves given by the present theory with 
$q = 0.391$ $(\mu = 0.240)$.
For the theoretical curves, the number of steps in the cascade $n$ are,
from top to bottom: 21.5, 20.0, 16.8, 14.0, 11.8, 10.1, 9.30, 8.10, 7.00 and 6.00.
For better visibility, each PDF is shifted by $-1$ unit along the vertical axis.
}
\label{experiment by Gotoh 381L log}
\end{center}
\end{figure}

There usually appears an asymmetry (a negative skewness) 
in the PDF's of the {\it longitudinal} 
velocity fluctuations, but this is not so vivid in the PDF's of the {\it transverse}
ones (see, for example, Fig.~15 and Fig.~16 in \cite{Gotoh01}). 
%This may be understood in the following way.
%For the longitudinal velocity fluctuation $\delta u_n^{(-)}$ 
%($\delta u_n^{(+)}$) with $u(x+r)-u(x)<0$ ($u(x'+r)-u(x')>0$),
%the velocity gradient is apt to take relatively larger negative (smaller positive) 
%values there in average. 
%Therefore, at a given separation $r \sim \ell_n >0$, we have
%$\delta u_n^{(-)} > \delta u_n^{(+)}$.
%The relative portion between the positive and the negative slope 
%of velocity gradients in physical space should be determined according to 
%a balance with pressure gradient.
%If the probability $\itPi(-\delta u_n^{(-)}/\delta u_0)$ and
%$\itPi(\delta u_n^{(+)}/\delta u_0)$ are about the same, the PDF may have
%a negative skewness.
Assuming that the singularities in the velocity gradient may contribute mainly to
the symmetric part, we will deal with the symmetric part of the 
PDF's of velocity fluctuations in the following.

With the help of the new variable 
\be
\xi_n = \delta u_n / \dbra \delta u_n^2 \dket^{1/2} 
= x_n / \dbra x_n^2 \dket^{1/2} = \bar{\xi}_n \delta_n^{\alpha /3 -\zeta_2 /2},
\ee
scaled by the variance of velocity fluctuations,
the PDF $\hat{\itPi}^{(n)}(\vert \xi_n \vert)$, introduced through
$
\hat{\itPi}^{(n)}(\vert \xi_n \vert) d\xi_n 
= \itPi^{(n)}(\vert x_n \vert) dx_n,
$
is given by \cite{AA5,AA6}
\bea
\hat{\itPi}^{(n)}(\xi_n) \aeq \hat{\itPi}_{<*}^{(n)}(\xi_n)
\quad \mbox{for $\vert \xi_n \vert \leq \xi_n^*$}
\\
\hat{\itPi}^{(n)}(\xi_n) \aeq \hat{\itPi}_{*<}^{(n)}(\xi_n)
\quad \mbox{for $\xi_n^* \leq \vert \xi_n \vert \leq 
\bar{\xi}_n \delta_n^{\alpha_{\rm min} /3 -\zeta_2 /2}$}.
\eea
Here, 
$
\bar{\xi}_n = [2 \gamma_2^{(n)} \delta_n^{-\zeta_2} + (1-2\gamma_0^{(n)} ) 
a_2 ]^{-1/2}
$.
Assuming that, for smaller velocity fluctuations $\vert \xi_n \vert \lesssim \xi_n^*$, 
the contribution to the PDF of the velocity fluctuations comes mainly from
thermal fluctuations related to the kinematic viscosity, 
we take for the PDF $\hat{\itPi}_{<*}^{(n)}(\xi_n)$ a Gaussian function \cite{AA5,AA6}, i.e.,
\be
\hat{\itPi}_{<*}^{(n)}(\xi_n) = \bar{\itPi}_{\rS}^{(n)}
\me^{-[1+3f'(\alpha^*)] [(\xi_n/\xi_n^* )^2 -1 ] /2}
\label{PDF less}
\ee
with
$
\bar{\itPi}_{\rS}^{(n)} = 3 (1-2\gamma_0^{(n)} )
/ (2 \bar{\xi}_n \sqrt{2\pi X \vert \ln \delta_n \vert} )
$.
On the other hand, we assume that the main contribution to 
$\hat{\itPi}_{*<}^{(n)}(\xi_n)$ may come from 
the multifractal distribution of singularities~\cite{AA4,AA5,AA6} related to
the turbulent viscosity, i.e.,
$
\hat{\itPi}_{*<}^{(n)}(\xi_n) = \hat{\itPi}_{\rS}^{(n)}(\vert \xi_n \vert)
$:
\be
\hat{\itPi}_{*<}^{(n)}(\xi_n)
= \bar{\itPi}_{\rS}^{(n)}\frac{\bar{\xi}_n}{\vert \xi_n \vert}
\left[1 - \left(\frac{3 \ln \left\vert \xi_n / \xi_{n,0} \right\vert}
{\Delta \alpha \ \vert \ln \delta_n \vert} \right)^2
\right]^{n/(1-q)}
\label{PDF larger}
\ee
with 
$
\vert \xi_{n,0} \vert = \bar{\xi}_n \delta_n^{\alpha_0 /3 -\zeta_2 /2}
$.
The point $\xi_n^*$ was defined by 
$
\xi_n^* = \bar{\xi}_n \delta_n^{\alpha^* /3 -\zeta_2 /2}
$
where $\alpha^*$ is the solution of 
$
\zeta_2/2 -\alpha/3 +1 -f(\alpha) = 0
$,
and $\hat{\itPi}_{<*}^{(n)}(\xi_n)$ and $\hat{\itPi}_{*<}^{(n)}(\xi_n)$ 
were connected at $\xi_n^*$ under the condition that they should have 
the same value and the same derivative there.
Here, 
\be
f(\alpha) = 1 + (1-q)^{-1} \log_2 \left[ 1 - \left(\alpha - \alpha_0\right)^2
\big/ \left(\Delta \alpha \right)^2 \right]
\label{Tsallis f-alpha}
\ee
is the multifractal spectrum~\cite{AA1,AA2,AA3,AA4}, derived by 
the relation
$
P^{(n)}(\alpha) \propto \delta_n^{1-f(\alpha)}
$ \cite{Meneveau87b,AA4},
that reveals how densely each singularity, 
labeled by $\alpha$, fills physical space.

Now, we proceed to analyze the data of DNS conducted by 
Gotoh et al.\ \cite{Gotoh01} for fully developed turbulence.
In the following theoretical analyses, we will take $\delta = 2$ for 
the number of pieces of "eddies" generated at each step in the energy cascade.
In Fig.~\ref{Fig: L scaling exponent} and Fig.~\ref{Fig: T scaling exponent}, 
we put, respectively, the measured scaling exponents $\zeta_m^{\rm L}$ for longitudinal 
velocity fluctuations and $\zeta_m^{\rm T}$ for transverse ones 
at $\rR_\lambda = 381$ (closed circles) \cite{Gotoh01}.
The scaling exponents (\ref{zeta}) derived by the present theory are
given by a solid line in each figure.
There are also represented, as references, the predictions of K41 
(dotted line) \cite{K41},  of She-Leveque (dashed line) \cite{She94}
and of the log-normal model (short-dashed line) \cite{Oboukhov62,K62,Yaglom}.
We determine, self-consistently, the values of the intermittency exponent $\mu$ 
by fitting the formula (\ref{zeta}) with the ten observed values of the scaling exponents, 
$\zeta_m^{\rm L}$ and $\zeta_m^{\rm T}$, with the help of the method of least squares.
The determined values of $\mu$ are listed in Table~\ref{parameters} with the values of 
the corresponding parameters $q$, $\alpha_0$ and $X$.
The latter three parameters are obtained by the $\mu$-dependent functions that
are given in this paper as the solutions of the self-consistent equations.
Note that the theoretical lines thus determined locate within the experimental 
error bars at each ten observed points. 
Therefore, the relation $\mu = 2 - \zeta_6$ is satisfied within 
the experimental error bars. 
We obtain
$
\alpha_{+} -\alpha_0 
= \alpha_0 - \alpha_{-} = 0.6818
$ (0.8167),
$
\itDelta \alpha = 1.160
$ (1.566)
for the longitudinal (transverse) fluctuations.

There is an argument that the scaling exponents $\zeta_2^{\rm L}$ 
for the longitudinal velocity structure function and $\zeta_2^{\rm T}$
for the transverse function should be equal for the isotropic and incompressible
turbulence. In Table~\ref{parameters}, we see that $\zeta_2^{\rm L} = 0.696$
and $\zeta_2^{\rm T} = 0.707$ within the present theoretical analysis, which 
give, respectively, 1.696 and 1.707 for the exponent of 
the Kolmogorov spectrum which is $5/3 = 1.\dot{6}$ in K41 \cite{K41}.
The small deviation (1\%) between $\zeta_2^{\rm L}$ and $\zeta_2^{\rm T}$ 
can be attributed to the finite sample size 
and the small amount of flow anisotropy \cite{Gotoh01}.
The scaling exponent $\zeta_4$ of the fourth order moment, related to 
the second order pressure structure function, is reported in \cite{Gotoh01} as
$\zeta_4^{\rm L} = 1.30 \pm 0.02$ and $\zeta_4^{\rm T} = 1.26 \pm 0.03$ 
at $R_\lambda = 381$.
These values are comparable with the ones in Table~\ref{parameters}, i.e.,
$\zeta_4^{\rm L} = 1.277$ and $\zeta_4^{\rm T} = 1.256$ derived by
the present theory.

\begin{figure}[thbp]
\begin{center}
\leavevmode
\epsfxsize=70mm
\epsfbox{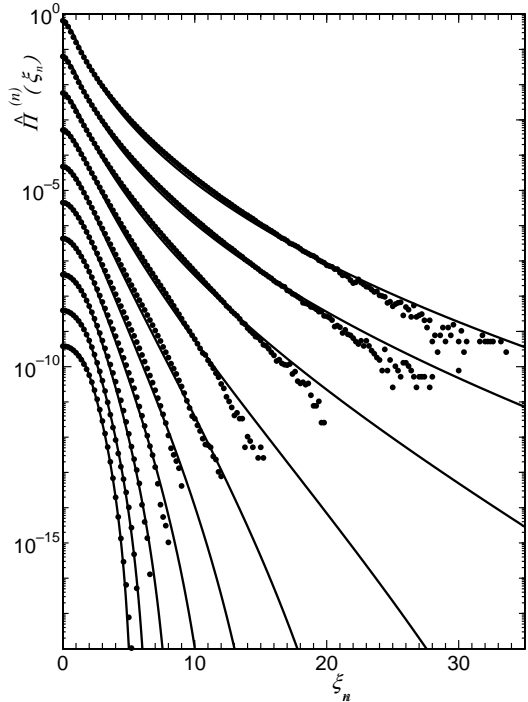}
\caption{Experimentally measured PDF of the {\it transverse} velocity fluctuations by 
Gotoh et al.\ for $R_\lambda =381$ 
are compared with the present theoretical results $\hat{\itPi}^{(n)}(\xi_n)$.
Closed circles are the symmetrized points. 
The separations $r/\eta = \ell_n/\eta$ are the same as 
in Fig.~\ref{experiment by Gotoh 381L log}. 
Solid lines represent the curves given by the present theory with 
$q = 0.543$ $(\mu = 0.327)$.
The number of steps in the cascade $n$ are,
from top to bottom: 20.0, 18.2, 14.6, 11.4, 9.30, 7.90, 6.80, 5.60, 4.70 and 4.00.
%For better visibility, each PDF is shifted by -1 unit along the vertical axis.
}
\label{experiment by Gotoh 381T log}
\end{center}
\end{figure}

The comparison between measured PDF's of the velocity fluctuations 
at $R_\lambda = 381$ by DNS \cite{Gotoh01} and 
those obtained by the present analysis \cite{AA4,AA5,AA6}
is given in Fig.~\ref{experiment by Gotoh 381L log} for longitudinal fluctuations
and in Fig.~\ref{experiment by Gotoh 381T log} for transverse ones.
In order to extract the symmetrical part from the DNS data, we took
mean average of those on the left hand side and on the right hand side.
The symmetrized data are described by closed circles.
The solid lines are the curves $\hat{\itPi}^{(n)}(\xi_n)$ given by 
(\ref{PDF less}) and (\ref{PDF larger}) with 
the values of parameters in Table~\ref{parameters}.
The values of $\xi_n^*$ are about 1 to 1.5.
The separations $r/\eta = \ell_n/\eta$ are, 
from top to bottom: 2.38, 4.76, 9.52, 19.0, 38.1, 76.2, 152, 305, 609 and 1220
for DNS \cite{Gotoh01} data.
On the other hand, the number $n$ of steps in the cascade for 
longitudinal (transverse) fluctuations are, from top to bottom: 
21.5, 20.0, 16.8, 14.0, 11.8, 10.1, 9.30, 8.10, 7.00 and 6.00
(20.0, 18.2, 14.6, 11.4, 9.30, 7.90, 6.80, 5.60, 4.70 and 4.00).
These values are obtained by the method of least squares with respect to
the logarithm of PDF's for the best fit of our theoretical formulae 
(\ref{PDF less}) and (\ref{PDF larger}) to the observed values of the PDF's 
by discarding those points which have observed values less than 10$^{-9}$ (10$^{-8}$)
since they scatter largely in the logarithmic scale. 
We see an excellent agreement between the measured PDF's and 
the analytical formula of PDF derived by the present self-consistent theory.

\begin{figure}[htbp]
\begin{center}
\leavevmode
\epsfxsize=100mm
\epsfbox{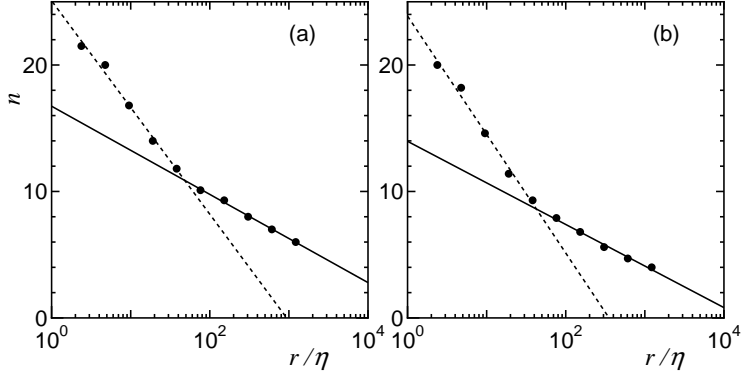}
\caption{Dependence of $n$ on $r/\eta$ is given by closed circles. 
(a) Those extracted from Fig.~\ref{experiment by Gotoh 381L log} 
for {\it longitudinal} fluctuations. The crossover occurs at $\ell_c^{\rm L}/\eta = 48.26$.
(b) Those from Fig.~\ref{experiment by Gotoh 381T log} for {\it transverse} fluctuations.
The crossover occurs at $\ell_c^{\rm T}/\eta = 42.57$.
}
\label{eoeta-n by Gotoh 381}
\end{center}
\end{figure}

The dependence of $n$ on $r/\eta$ for longitudinal (transverse) fluctuations,
extracted from Fig.~\ref{experiment by Gotoh 381L log} 
(Fig.~\ref{experiment by Gotoh 381T log}), 
is shown in Fig.~\ref{eoeta-n by Gotoh 381} (a) 
(Fig.~\ref{eoeta-n by Gotoh 381} (b)) by solid and dashed lines.
These lines are obtained by the method of least squares within linear fits,
and are given by
\bea
n \aeq -1.050 \times \log_2 r/\eta + 16.74
\quad (\mbox{for } \ell_c^{\rm L} \leq r),
\label{n-roeta L larger} \\
n \aeq -2.540 \times \log_2 r/\eta + 25.08
\quad (\mbox{for } r < \ell_c^{\rm L})
\label{n-roeta L less}
\eea
with the crossover length $\ell_c^{\rm L}/\eta = 48.26$ 
for longitudinal fluctuations, and
\bea
n \aeq -0.9896 \times \log_2 r/\eta + 13.95
\quad (\mbox{for } \ell_c^{\rm T} \leq r),
\label{n-roeta T larger} \\
n \aeq -2.820 \times \log_2 r/\eta + 23.87
\quad (\mbox{for } r < \ell_c^{\rm T})
\label{n-roeta T less}
\eea
with the crossover length $\ell_c^{\rm T}/\eta = 42.57$ 
for transverse fluctuations.
We see that $\ell_c^{\rm T}/\eta \lesssim \ell_c^{\rm L}/\eta$, and that
these crossover lengths have values close to the reported value of 
the Taylor microscale $\lambda /\eta = 38.33$ in \cite{Gotoh01} at $R_\lambda = 381$.
The value of $\ell_c^{\rm T}/\eta$ is very close to the one
in Fig.~31 of \cite{Gotoh01}.
Whereas, the values of $\ell_c^{\rm L}/\eta$ is different from the one obtained 
in \cite{Gotoh01} where Gotoh et al.\ got the relation 
$\ell_c^{\rm L}/\eta \approx 2\ell_c^{\rm T}/\eta$ within their analyses.
In spite of the difference in the values of $\ell_c^{\rm L}/\eta$,
the inequality $\ell_c^{\rm T}/\eta < \ell_c^{\rm L}/\eta$ 
may be attributed to a manifestation of the incompressible nature
of the fluid under consideration as interpreted in \cite{Gotoh01}.
Notice that this kind of crossover was not observed in the analysis \cite{AA5} 
of the experiment conducted by Lewis and Swinney \cite{Lewis-Swinney99} for 
turbulent Couette-Taylor flow.
%These properties are consistent with those found in \cite{Gotoh01}.

The equations (\ref{n-roeta L larger}) and (\ref{n-roeta T larger}) have 
the slope close to $-1$ with respect to $\log_2 r/\eta$ that is consistent 
with (\ref{n-roeta no intermittency}) for $\delta = 2$. 
Therefore, we conclude that the {\it upper} scaling region 
$\ell_c^{\rm L}/\eta \leq r/\eta \lesssim 1220$ 
($\ell_c^{\rm T}/\eta \leq r/\eta \lesssim 1220$) 
found within the present systematic analysis 
can be interpreted as the inertial range for longitudinal (transverse) fluctuations.
Note that this inertial range for longitudinal fluctuations
is wider than the longitudinal scaling range 
$2\lambda/\eta \lesssim r/\eta \lesssim 220.7$ observed in \cite{Gotoh01} 
at $R_\lambda = 381$. 
The latter estimate was done by looking for the flat region of
the observed $\zeta_m^{\rm L}$ with respect to the separation $r/\eta$.
The existence of the crossover length was observed first by Gotoh et al.\ 
in the analyses of their DNS data \cite{Gotoh01} 
as the low end of the scaling range for the longitudinal or the transverse fluctuations, 
and was attributed to the existence of structure of this order in turbulence, 
e.g., a shear layer or a vortex tube.

It is remarkable that in Fig.~\ref{eoeta-n by Gotoh 381} we can see,
within the present self-consistent analysis, the existence of 
another new scaling region, the {\it lower} scaling region, i.e., 
$2 \lesssim r/\eta \leq \ell_c^{\rm L}/\eta$ 
($2 \lesssim r/\eta \leq \ell_c^{\rm T}/\eta$) 
for longitudinal (transverse) velocity fluctuations.
It may be interesting to point out that the slope in (\ref{n-roeta L less}) 
and (\ref{n-roeta T less}) can be $-1$ with respect to $\log_\delta r/\eta$ 
with $\delta = \delta^{\rm L} = 1.33$ for longitudinal fluctuations 
and with $\delta = \delta^{\rm T} = 1.28$ for transverse ones, respectively. 
Therefore, in this {\it lower} scaling region, it can be interpreted that 
eddies break up, effectively, into $\delta^{\rm L}$ ($\delta^{\rm T}$) pieces 
at each step of the energy cascade for longitudinal (transverse) fluctuations.

Assuming that (\ref{n-roeta no intermittency}) with $\delta = 2$ 
is applicable in the "inertial range" 
$\ell_c^{\rm T}/\eta, \ell_c^{\rm L}/\eta \leq r/\eta \lesssim \ell_0/\eta$,
we can extract from (\ref{n-roeta L larger}) and (\ref{n-roeta T larger}) 
the values of the Reynolds number $\rRe^{\rm L} = 5.236 \times 10^{6}$ 
for longitudinal eddies, and of 
$\rRe^{\rm T}= 7.113 \times 10^{5}$ for transverse eddies.
If we adopt the relation
$
R_\lambda^2 = A^{\rm L} \rRe^{\rm L} = A^{\rm T} \rRe^{\rm T}
$ \cite{Batchelor53},
we have $A^{\rm L}= 0.02772$ and $A^{\rm T}= 0.2041$.

In this paper, we showed how precisely the formulae, derived by our theory 
\cite{AA1,AA2,AA3,AA4,AA5,AA6}, can explain the data observed in the DNS \cite{Gotoh01}.
We found, explicitly, that there exist two scaling regions, i.e., 
the {\it upper} scaling region with larger separations which may correspond 
to the scaling range observed by Gotoh et al.,
and the {\it lower} scaling region with smaller separations which is 
a new scaling region extracted first by the present systematic analysis.
These scaling regions are divided by a definite length approximately of 
the order of the Taylor microscale $\lambda$, that may correspond to 
the crossover length introduced by Gotoh et al.\ \cite{Gotoh01} as the length 
at which a scaling behavior of the structure functions ceases due to 
the effect of kinematic viscosity. 
We also saw that the inertial range (the {\it upper} scaling region) 
for longitudinal fluctuations is wider than the scaling range 
extracted by Gotoh et al.\ \cite{Gotoh01}.
The discovery of the {\it lower} scaling region by the present analysis may tell us 
that the distribution of singularities of the velocity gradient in physical space 
is robust enough to produce a scaling behavior even in the dissipation range.
In other words, the system of turbulence may have an intrinsic scaling property,
applicable in a wider region than the inertial range, whose origin can be
attributed to the multifractal distribution of the singularities in 
the velocity gradient.
A study of the PDF's for velocity derivatives by the present analysis 
is one of the attractive open problems and is now in progress.
It is also one of the attractive future problems to provide with a dynamical approach
based on the aspect given by the present ensemble theoretical approach 
in order to interpret, especially, the {\it lower} scaling region. 
These will be reported elsewhere.

%We would like to close this paper by mentioning something about the standard average
%and the $q$-average. 
%We interpret that any observed value should be given by a standard average of 
%corresponding observable quantity.
%The $q$-average was used just in the derivation of the distribution 
%(\ref{Tsallis prob density}) of $\alpha$ which is not considered as 
%an observable quantity.

The authors are grateful to Prof.~T.~Gotoh for enlightening discussion
and his kindness to show his data prior to publication.
The authors would like to thank Prof.~C.~Tsallis and Dr.~A.K.~Rajagopal 
for their fruitful comments with encouragement.

%%%%%%%%%%%%%%%%%%%%%%%%%%%%%%%%%%%%%%%%%%%%%%%%%%%%%%%%%%%%%%%%%
% References:
%%%%%%%%%%%%%%%%%%%%%%%%%%%%%%%%%%%%%%%%%%%%%%%%%%%%%%%%%%%%%%%%%

\begin{table}[tbp]
\begin{center}
\begin{tabular}{ccccccc}
  & \multicolumn{2}{c}{longitudinal} & \multicolumn{2}{c}{transverse} \\
\hline\hline
$\mu$      & \multicolumn{2}{c}{0.240}  & \multicolumn{2}{c}{0.327} \\
\hline
$q$        & \multicolumn{2}{c}{0.391}  & \multicolumn{2}{c}{0.543} \\
$\alpha_0$ & \multicolumn{2}{c}{1.138}  & \multicolumn{2}{c}{1.189} \\
$X$        & \multicolumn{2}{c}{0.285}  & \multicolumn{2}{c}{0.388} \\
\hline\hline \\[-5pt]
$m$  & \multicolumn{2}{c}{$\zeta_m^{\rm L}$}  & \multicolumn{2}{c}{$\zeta_m^{\rm T}$} \\ 
   & present theory    & DNS & present theory &  DNS  \\
\hline
1  & 0.3637  & 0.370$\pm$0.004 & 0.3747 & 0.369$\pm$0.004 \\ 
2  & 0.6965  & 0.709$\pm$0.009 & 0.7073 & 0.701$\pm$0.01  \\ 
3  & 1.000  & 1.02$\pm$0.02    & 1.000 & 0.998$\pm$0.02  \\ 
4  & 1.277   & 1.30$\pm$0.02   & 1.256  & 1.26$\pm$0.03    \\
5  & 1.529   & 1.56$\pm$0.03   & 1.480  & 1.49$\pm$0.04   \\
6  & 1.761   & 1.79$\pm$0.04   & 1.674  & 1.69$\pm$0.05   \\
7  & 1.973   & 1.99$\pm$0.04   & 1.843  & 1.86$\pm$0.05   \\
8  & 2.169   & 2.18$\pm$0.04   & 1.990  & 2.00$\pm$0.04   \\
9  & 2.350   & 2.35$\pm$0.04   & 2.117  & 2.11$\pm$0.05   \\
10 & 2.519   & 2.49$\pm$0.04   & 2.227  & 2.20$\pm$0.06    
\end{tabular}
\end{center}
\caption{Comparison of the values of the scaling exponents (\ref{zeta}) derived by 
the present theory and those for the {\it longitudinal} and 
{\it transverse} velocity fluctuations 
observed in DNS conducted by Gotoh et al.. The value $\mu$ for each fluctuations
is also listed with the corresponding values for $q$, $\alpha_0$ and $X$.
}
\label{parameters}
\end{table}


\begin{references}

\bibitem{K41} A.N.~Kolmogorov, 
		C.R.~Acad. Sci. USSR {\bf 30}, 301; 538 (1941).
\bibitem{Heisenberg48} W.~Heisenberg, Z.~Phys. {\bf 124}, 628 (1948).
\bibitem{Landau59} L.D.~Landau and E.M.~Lifshitz, {\it Fluid Mechanics} 
		(Addison-Wesley, 1959).
\bibitem{Oboukhov62} A.M.~Oboukhov, J. Fluid Mech. {\bf 13},
		77 (1962).
\bibitem{K62} A.N.~Kolmogorov, J. Fluid Mech. {\bf 13},
		82 (1962).
\bibitem{Yaglom} A.M.~Yaglom, Sov. Phys. Dokl. {\bf 11},
		26 (1966).
\bibitem{Frisch78} U.~Frisch, P-L.~Sulem and M.~Nelkin,
		J. Fluid Mech. {\bf 87}, 719 (1978).
\bibitem{Meneveau87a} C.~Meneveau and K.~R.~Sreenivasan,
		Phys. Rev. Lett. {\bf 59}, 1424 (1987).
\bibitem{Meneveau87b} C.~Meneveau and K.R.~Sreenivasan,
		Nucl. Phys. B (Proc. Suppl.) {\bf 2}, 49 (1987).
\bibitem{Hosokawa91} I.~Hosokawa, Phys. Rev. Lett. {\bf 66}, 1054 (1991).
\bibitem{Renyi} A.~R\'{e}nyi, {\it Proc.\ 4th Berkeley Symp.\ 
		Maths.\ Stat.\ Prob.} {\bf 1}, 547 (1961).
\bibitem{Tsallis88} C.~Tsallis, J. Stat. Phys. {\bf 52}, 479 (1988).
\bibitem{Tsallis99} C.~Tsallis, Braz. J. Phys. {\bf 29}, 1 (1999);
	On the related recent progresses see at\\ http://tsallis.cat.cbpf.br/biblio.htm.
\bibitem{AA} T.~Arimitsu and N.~Arimitsu, Phys. Rev. E {\bf 61},
		 3237 (2000).
\bibitem{AA1} T.~Arimitsu and N.~Arimitsu, J. Phys. A: Math. Gen. {\bf 33}, 
		L235 (2000)  [{\footnotesize CORRIGENDUM}: {\bf 34}, 673 (2001)].
\bibitem{AA2} T.~Arimitsu and N.~Arimitsu, Chaos, Solitons and Fractals 
		{\bf 13}, 479 (2002).
\bibitem{AA3} T.~Arimitsu and N.~Arimitsu, Prog.~Theor.~Phys. {\bf 105}, 
		 355 (2001).
\bibitem{AA4} T.~Arimitsu and N.~Arimitsu, Physica A {\bf 295}, 177 (2001).
\bibitem{AA5} N.~Arimitsu and T.~Arimitsu, cond-mat/0109132 (2001).
\bibitem{AA6} T.~Arimitsu and N.~Arimitsu, cond-mat/0109007 (2001).
\bibitem{Gotoh01} T.~Gotoh, D.~Fukayama and T.~Nakano (2001) preprint.
\bibitem{Lewis-Swinney99} G.S.~Lewis and H.L.~Swinney, Phys. Rev. E {\bf 59}, 5457 (1999).
\bibitem{Beck-Lewis-Swinney01} C.~Beck, G.S.~Lewis and H.L.~Swinney, Phys. Rev. 
		E {\bf 63}, 035303-1 (2001).
\bibitem{Batchelor53} G.K.~Batchelor, {\it The Homogeneous Turbulence}
		(Cambridge Univ. Press., Cambridge, 1953).
\bibitem{Frisch-Parisi83} U.~Frisch and G.~Parisi, in {\it Turbulence
		and Predictability in Geophysical Fluid Dynamics and Climate 
		Dynamics}, ed.\ by M.~Ghil, R.~Benzi and G.~Parisi (North-Holland,
		New York, 1985) 84.
\bibitem{Benzi84} R.~Benzi, G.~Paladin, G.~Parisi and A.~Vulpiani, 
		J. Phys. A: Math. Gen. {\bf 17}, 3521 (1984).
\bibitem{Havrda-Charvat} J.H.~Havrda and F.~Charvat,
		Kybernatica {\bf 3}, 30 (1967).
\bibitem{Costa} U.M.S.~Costa, M.L.~Lyra, A.R.~Plastino and
		C.~Tsallis, Phys. Rev. E {\bf 56}, 245 (1997).
\bibitem{Lyra98} M.L.~Lyra and C.~Tsallis, 
		Phys. Rev. Lett. {\bf 80}, 53 (1998).
\bibitem{She94} Z-S.~She and E.~Leveque, Phys. Rev. Lett. 
		{\bf 72}, 336 (1994).
%\bibitem{AA01Cagliari} T.~Arimitsu and N.~Arimitsu, Lecture given in the {\it International
%		School and Workshop on Non Extensive Thermodynamics and Physical Applications}
%		at Cagliari in Italy in May 2001.

\end{references}
\end{document}